\documentclass[fleqn,10pt]{wlscirep}
\usepackage[utf8]{inputenc}
\usepackage[T1]{fontenc}
\usepackage{amsmath}
\usepackage[outdir=./]{epstopdf}
\title{Time-scale dependence of solar wind-based regression models of ionospheric electrodynamics}

\author[1,*]{Karl M. Laundal}
\author[1]{Jone P. Reistad}
\author[1]{Spencer M. Hatch}
\author[1]{Therese Moretto}
\author[1]{Anders Ohma}
\author[1]{Nikolai \O{}stgaard}
\author[1]{Paul A. R. Tenfjord}
\author[2]{Christopher C. Finlay}
\author[2]{Clemens Kloss}
\affil[1]{Birkeland Centre for Space Science, Department of Physics and Technology, University of Bergen, Bergen, 5006, Norway}
\affil[2]{DTU Space, National Space InstituteTechnical University of Denmark, Kgs. Lyngby, Denmark}

\affil[*]{karl.laundal@uib.no}

\begin{abstract}
The solar wind influence on geospace can be described as the sum of a directly driven component, or dayside reconnection, and an unloading component, associated with the release of magnetic energy via nightside reconnection. The two processes are poorly correlated on short time scales, but exactly equal when averaged over long time windows. Because of this peculiar property, regression models of ionospheric electrodynamics that are based on solar wind data are time scale specific: Models derived from 1~min resolution data will be different from models derived from hourly, daily, or monthly data. We explain and quantify this effect on simple linear regression models of various geomagnetic indices. We also derive a time scale-dependent correction factor that can be used with the Average Magnetic field and Polar current System model. Finally, we show how absolute estimates of the nightside reconnection rate can be calculated from solar wind measurements and geomagnetic indices.
\end{abstract}
\begin{document}

\flushbottom
\maketitle
\thispagestyle{empty}

\section*{Introduction}

The solar wind carries the energy that shapes the magnetosphere and powers auroras, plasma flows, and electric currents at high latitudes. Because the key controlling solar wind parameters $-$ the speed, density, and the magnetic field that the solar wind carries with it $-$ have been reliably measured at L1 for several decades, many empirical models of ionospheric electrodynamics \cite{Weimer05,Weimer13,Laundal18b} are parametrized in terms of these measurements. 

This practice, however, ignores the large variations in time scales of the solar wind influence on geospace: Magnetic reconnection between the interplanetary and terrestrial magnetic fields on the dayside leads to changes in flows and currents in the ionosphere typically within less than 20~min \cite{Snekvik17}. Simultaneously, solar wind kinetic energy is converted to magnetic energy that builds up in the magnetotail lobes. Sometime later, typically hours, this energy is released through nightside reconnection, and flows and currents are again excited in the ionosphere. Solar wind measurements are good indicators of the first of these two steps, but much less useful in predicting when the magnetotail energy conversion will take place. On the other hand, since the nightside and dayside reconnection rates on average must balance, solar wind measurements provide an excellent indication of both nightside and dayside processes on long time scales. In this paper, we discuss the effects of this paradoxical time scale property on solar wind-based regression models of ionospheric electrodynamics. 

As a starting point, we make use of the so-called expanding contracting polar cap paradigm \cite{Siscoe85,Lockwood91,Cowley92,Milan07}, which explains how the excitation and decay of ionospheric flows are related to nightside and dayside reconnection. A central result of this paradigm \cite{Lockwood91} is that, for a circular polar cap, the cross polar cap potential $V$ is given by
\begin{equation}
V = (\Phi_D + \Phi_N)/2 \label{eq:CPCP}
\end{equation}
where $\Phi_D$ and $\Phi_N$ are the dayside and nightside reconnection rates. It is clear from this equation that any statistical model of $V$, or quantitites correlated with $V$, that depends on $\Phi_D$ but not $\Phi_N$ will be imperfect. 

The effects of $\Phi_N$ may be at least partially present even in models that are based only on estimates of $\Phi_D$. Consider the following approximate model for $\Phi_N$ in terms of $\Phi_D$:
\begin{equation}
\Phi_N^\tau = c(\tau) + d(\tau)\Phi_D^\tau \label{eq:phin_vs_phid}
\end{equation}
where the superscripts $\tau$ denote an averaging time window, here also referred to as \emph{time scale}, defined as follows:
\begin{equation}
y^\tau(t) = \frac{1}{\tau}\int_{t - \tau}^ty(t)dt. \label{eq:time_averaging}
\end{equation}
The parameters in equation (\ref{eq:phin_vs_phid}) are functions of $\tau$ because of the two-step response to solar wind driving discussed above. When $\tau$ is small (i.e., of the order of minutes or less), $d(\tau)$ is correspondingly small since $\Phi_D$ and $\Phi_N$ are not closely related on such short time scales. On the other hand, in the limit $\tau \rightarrow \infty$, $c(\infty) = 0$ and $d(\infty) = 1$ since dayside and nightside reconnection rates must balance (i.e., $\Phi_D^\infty = \Phi_N^\infty$).

Because of the time scale dependent relationship between $\Phi_D$ and $\Phi_N$, statistical models of some measurable quantity related to magnetosphere/ionosphere convection that use $\Phi_D$ are also dependent on time scales. To quantify this dependence, assume that such a measurable quantity, $y$, depends on $V$ as follows:
\begin{equation}
y = \alpha + \beta V. \label{eq:ideal_model}
\end{equation}
$y$ is here a generic term for any parameter that follows equation (\ref{eq:ideal_model}), and will later be replaced by specific geomagnetic indices. The physical justification for equation (\ref{eq:ideal_model}) is discussed further in the next section. This equation is assumed to be valid on all time scales, and $\alpha$ and $\beta$ do not depend on $\tau$. Since we do not know $V$, equation (\ref{eq:ideal_model}) is not very useful for making empirical models of $y$. Instead, it is common to use a solar wind magnetosphere coupling function \cite{Newell07,Milan12,Tenfjord13}, $\epsilon$. Such coupling functions combine solar wind measurements in different ways to maximize the correlation with the transfer of energy or magnetic flux from the solar wind to the magnetosphere. We will assume that they correlate with the transfer of mganetic flux, i.e., the dayside reconnection rate $\Phi_D$:
\begin{equation}
\Phi_D = k\epsilon, \label{eq:epsilon}
\end{equation}
where $k$ is a proportionality constant that scales $\epsilon$ such that its amplitude matches that of the dayside reconnection rate. The unit of equation (\ref{eq:epsilon}) is magnetic flux per second [Wb/s], which is equal to volt [V]. The Milan coupling function \cite{Milan12} is a special case, where the function includes an empirically determined scale factor so that $k=1$. 

Using the above equations, we can derive a model of $y^\tau$ that depends only on $\epsilon^\tau$. The model will be time scale-dependent (hence the superscripts $\tau$), since we make use of equation (\ref{eq:phin_vs_phid}): Starting with the time-averaged equation (\ref{eq:ideal_model}), we express $V^\tau$ in terms of $\epsilon^\tau$ using equations (\ref{eq:CPCP}), (\ref{eq:phin_vs_phid}), and (\ref{eq:epsilon}):
\begin{align}
y^\tau &= \alpha + \beta V^\tau = \alpha + \beta (k\epsilon^\tau + c(\tau) + d(\tau)k\epsilon^\tau)/2, \nonumber\\
y^\tau &= a(\tau) + b(\tau) \epsilon^\tau \label{eq:y},
\end{align}
where
\begin{align}
a(\tau) &= \alpha + \beta c(\tau)/2 \label{eq:a_tau},\\ 
b(\tau) &= \beta k(1 + d(\tau))/2. \label{eq:b_tau}
\end{align}
Equation \ref{eq:y} is a linear model of $y^\tau$ in terms of $\epsilon^\tau$, with model parameters $a$ and $b$ that depend on $\tau$ since $c(\tau)$ and $d(\tau)$ from equation (\ref{eq:phin_vs_phid}) are time-scale dependent. The implication of this time scale dependence is that a regression model on this form which is derived using data with time resolution $\tau$ in general should not be compared with data having a different time resolution. In the next section we demonstrate this by estimating $a$ and $b$ for a range of time scales by replacing $y$ with specific geomagnetic indices. Later we discuss how the time-scale dependence of solar-wind based regression models affects the interpretation of more complex empirical models, using the Average Magnetic field and Polar current System (AMPS) model \cite{Laundal18b} as an example. The AMPS model is an empirical global model of ionospheric currents and associated magnetic field based on magnetic field measurements in low Earth orbit from the CHAMP and Swarm satellites. 

In the limit $\tau \rightarrow \infty$, equation (\ref{eq:a_tau}) reduces to $a(\infty) = \alpha$ since $c(\infty) = 0$, and equation (\ref{eq:b_tau}) reduces to $b(\infty) = \beta k$ since $d(\infty) = 1$. Using the above results, we can derive an expression for $\Phi_N$ in terms of $y$ and $\Phi_D$ that is independent of time scales: By solving equation (\ref{eq:CPCP}) for $\Phi_N$, and using (\ref{eq:ideal_model}) to replace $V$, we get
\begin{equation}
\Phi_N = 2(y-\alpha)/\beta - \Phi_D = 2(y-a(\infty))/b(\infty) - \Phi_D \label{eq:phin},
\end{equation}
where, in the last step, we used equations (\ref{eq:a_tau}) and (\ref{eq:b_tau}) in the limits $\tau \rightarrow \infty$ to replace $\alpha$ and $\beta$. As will be demonstrated in the next section, $a(\infty)$ and $b(\infty)$ can be estimated from data, by averaging over several days. If we use a coupling function that is scaled so that $k = \Phi_D/\epsilon = 1$, such as the function developed by \cite{Milan12}, $\Phi_N$ can be calculated from observations of $y$ and of the solar wind. Equation (\ref{eq:phin}) is in principle valid on any time scale of $\Phi_N$, $\Phi_D$, and $y$. We will later put this equation to use and discuss its limitations.

\section*{Time-scale dependence of solar wind based linear regression models of geomagnetic indices}\label{sec:regression}
Here we apply the equations from the previous section to a set of geomagnetic indices: AL, AU, PC, and ASY-H. These are four geomagnetic indices with 1~min time resolution, which are believed to describe various aspects of geomagnetic activity. The AL and AU indices monitor the westward and eastward auroral electrojets, respectively. The PC (polar cap) index is based on a magnetometer that is typically poleward of the auroral oval, and it is ostensibly an indirect measurement of the magnitude of the solar wind-magnetosphere coupling. The ASY-H index is a measure of the longitudinal variability of the magnetic perturbations at mid latitudes, and it tends to correlate with the auroral electrojet indices. For definitions and details about interpretation, see the review by Kauristie et al.\cite{Kauristie17}, and references therein. We will show that these parameters, when modeled as functions of a solar wind coupling function $\epsilon$, depend on the time scale $\tau$. 

For the sake of argument, we assume that each index is a linear function of dayside and nightside reconnection rates, and nothing else. In other words, they obey (\ref{eq:ideal_model}), but with different values for $\alpha$ and $\beta$. This is of course an approximation, since they would otherwise be exactly correlated. For example, we ignore conductivity effects for all indices except AU, which we scale by $(1 + \sin\psi)^{-1}$, where $\psi$ is the dipole tilt angle. This scaling is meant to roughly account for seasonal variations in solar EUV induced conductivity. The scaling improves the results with AU, but not with the other indices which may be more dependent on precipitation induced conductivity. In this analysis $\epsilon$ is the coupling function defined by Newell et al.\cite{Newell07}, but similar results are obtained when using other functions \cite{Tenfjord13,Milan12}. The Newell coupling function depends on the solar wind speed and the component of the interplanetary mganetic field (IMF) that is perpendicular to the Sun-Earth line. The solar wind data and indices are taken from the 1~min resolution OMNI database, between 1995 and 2018.

\begin{figure}
\includegraphics[width=.5\linewidth]{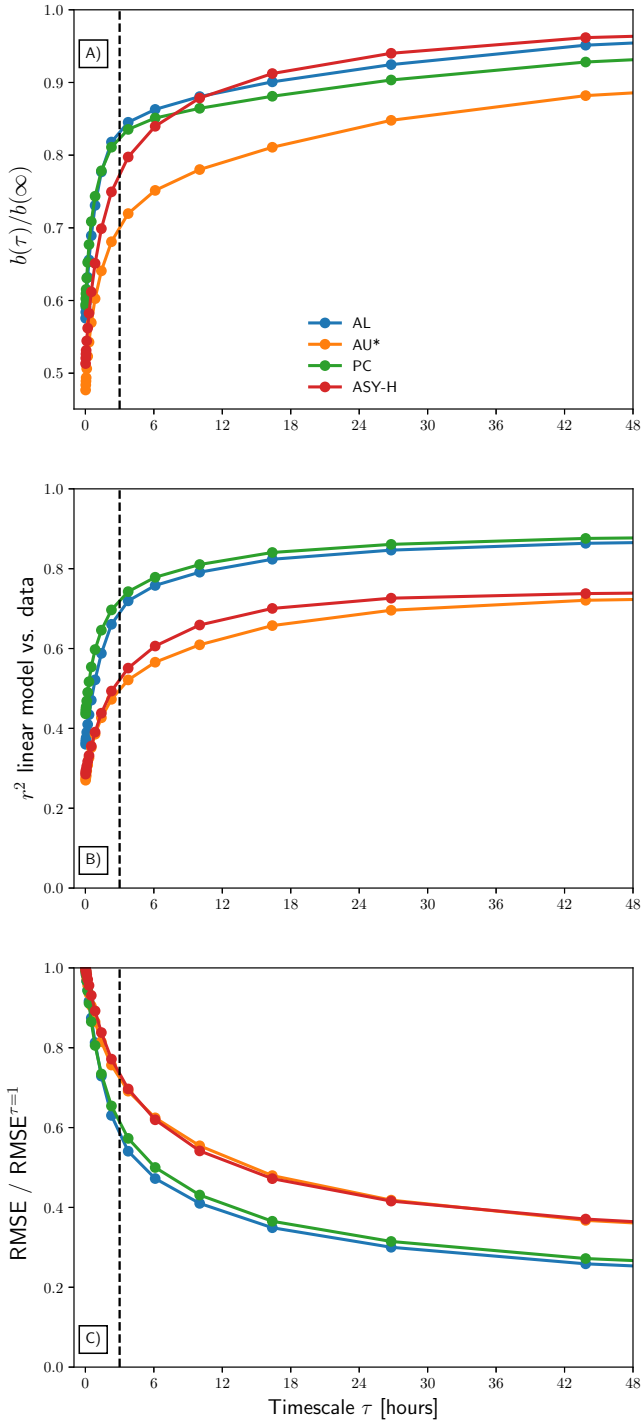}
\caption{Linear regression model statistics as functions of timescale $\tau$. Four models of the form $y^{\tau} = a(\tau) + b(\tau)\epsilon^\tau$ are considered, where $\epsilon$ is Newell's solar wind magnetosphere coupling function\cite{Newell07} and $y$ is the AL index, solar conductivity corrected AU index (AU*), the PC index, and the ASY-H index. The three statistics are A) the slope $b(\tau)$ normalized by $b(\infty)$, B) the Pearson correlation coefficient squared ($r^2$) between the model and the data, and C) the root-mean-square error (RMSE), normalized by the RMSE for $\tau = 1$~min, which is the time scale of the peak value. The vertical dashed lines mark $\tau = 3$~h, a typical substorm cycle period.}
\label{fig:param_model_stats}
\end{figure}

Figure \ref{fig:param_model_stats}A shows, for each of the four indices, the ratio $b(\tau)/b(\infty)$ where $b(\tau)$ is defined in equation (\ref{eq:y}) and estimated using ordinary least squares. $b(\infty)$ was calculated with $\tau = 8$~days. The ratio $b(\tau)/b(\infty)$ is related to the slope $d(\tau)$ in equation (\ref{eq:phin_vs_phid}), which describes the relationship between dayside and nightside reconnection rates on time scale $\tau$. Specifically, $d(\tau) = 2b(\tau)/b(\infty) - 1$, which can be found by solving equation (\ref{eq:b_tau}) for $d(\tau)$ and using that $\beta k = b(\infty)$. All curves in figure \ref{fig:param_model_stats}A follow a similar variation with $\tau$, with a steep increase at $\tau < 3$~h ($\tau=3$~h is indicated with a dashed bar), and then a more gradual increase towards 1. The transition at about 3~h is expected from previous studies \cite{Borovsky93,Freeman04} indicating that this is an average substorm cycle period (i.e., the time between bursts of tail reconnection associated with individual substorms).

Figure \ref{fig:param_model_stats}B shows the squared of the Pearson correlation coefficient $r^2$ between model and data. This coefficient is a measure of the fraction of variation in data that is explained by the model. For each index, $r^2$ is rapidly increasing for larger $\tau$. For the AL and PC indices, $r^2 > 0.8$ at large $\tau$. This indicates that (\ref{eq:ideal_model}) is indeed a good representation of these indices, and it reflects that $\epsilon = k\Phi_D = k\Phi_N$ as $\tau\rightarrow\infty$. Figure \ref{fig:param_model_stats}C shows the root mean square error (RMSE) of the model compared to the data, relative to the RMSE at $\tau = 1$~min. The basic result shown in Figure \ref{fig:param_model_stats} is that the models improve as $\tau$ increases; the physical reason for this trend is that $\epsilon$ becomes representative of both dayside and nightside reconnection. 

The indices studied here follow a skewed distribution. The peak of the distribution is near zero for each of the four indices that we consider. This represents a quiet day baseline. Active events tend to give negative excursions in the AL index and positive excursions in the other three indices. Since the indices do not follow a normal distribution, a quantitative interpretation of the Pearson correlation coefficient is questionable. However, in this study, the main purpose of the correlation coefficient is to qualitatively indicate how the association between our simple model and the observation that it describes vary with time scales. Given the very large number of measurements ($>20$~years of measurements at 1-min resolution), we believe this use is appropriate.

\section*{Effect of time-scale dependence on climatological models}\label{sec:amps}
Several climatological models of ionospheric electrodynamics depend on solar wind parameters, and lack parameters that represent the high-frequency portion of $\Phi_N$ \cite{Weimer13,Laundal18b}. The AMPS model \cite{Laundal18b} describes the ionospheric disturbance magnetic field in space, and the associated ionospheric currents. It is formulated in a way that makes it possible to write the model magnetic field as 
\begin{equation}
\mathbf{B} = \mathbf{B}_0(\dots) + \epsilon \mathbf{B}_\epsilon(\dots) \label{eq:amps_decomposition},
\end{equation}
where $\mathbf{B_0}$ and $\mathbf{B}_\epsilon$ are two different functions of space and of the external parameters used in the AMPS model (the IMF, solar wind velocity, dipole tilt angle and F$_{10.7}$ solar flux index). In the sum in equation (\ref{eq:amps_decomposition}), $\mathbf{B}_\epsilon$ is multiplied by Newell's solar wind-magnetosphere coupling function\cite{Newell07} $\epsilon$, and $\mathbf{B}_0$ is not. This is conceptually similar to the much simpler equation (\ref{eq:y}), and it implies that the AMPS model may also be time-scale dependent.

\subsection*{Relating magnetic field disturbances in space and ground} \label{sec:method}
To address the time scale dependence of the AMPS model, we compare model estimates with measurements of the magnetic field disturbances from ground magnetometers, instead of satellites. This is because time averaging satellite measurements mixes spatial and temporal effects. However, the AMPS model is based only on magnetic field measurements in space, above the conducting layer of the ionosphere where horizontal currents flow. This region is the only place where the full current system can be estimated using only magnetic field measurements \cite{Laundal16}. The reason for this is that below the ionosphere, the magnetic field is partially canceled. In this section we derive mathematical expressions which relate ground magnetic field perturbations to the AMPS model coefficients, thus allowing for calculation of model predictions of the AL index, which we will use in comparisons with measurements in the next section. 

We model the ionospheric current system as a two-dimensional sheet current on a sphere at some height $h_R$, set to 110~km here. The sheet current is connected to the magnetosphere by a vertical volume current that extends out from the sphere. At polar latitudes, where the Earth's magnetic field lines are almost vertical, such currents are approximately equal to the Birkeland currents (magnetic field-aligned electric currents). The sheet current can be decomposed further into divergence-free ($\mathbf{J}_{df}$) and curl-free ($\mathbf{J}_{cf}$) components. The latter is related to the radial current density $J_u$ by the current continuity equation $\nabla\cdot\mathbf{J}_{cf} = -J_u$. The magnetic fields of $\mathbf{J}_{cf}$ and $J_u$ cancel below the ionosphere, according to the Fukushima theorem \cite{Fukushima76}. Consequently, only the magnetic field of $\mathbf{J}_{df}$ must be considered. From now on we call that magnetic field $\mathbf{B}$, disregarding contributions from Birkeland/curl-free currents. We also assume that that contributions from the Earth's core, crust, and large-scale magnetosphere are accounted for by subtracting CHAOS-6 model predictions \cite{Finlay16}. The CHAOS-6 model is a high resolution, regularly updated, geomagnetic field model derived from magnetometer measurements from ground observatories and several satellites in low Earth orbit.

Above and below the current sheet, the magnetic field of $\mathbf{J}_{df}$ can be expressed as a gradient, $\mathbf{B} = -\nabla V$. In the AMPS model, $V$ is expressed as
\begin{equation}
V(\lambda_q, \phi_\mathrm{mlt}, h) = R_E\sum_{n,m}\left(\frac{R_E}{R_E + h}\right)^{n+1} P_n^m(\sin\lambda_q) [g_n^m\cos(m\phi_{\mathrm{mlt}}) + h_n^m\sin(m\phi_{\mathrm{mlt}})]. \label{eq:V_pol}
\end{equation}
where $\lambda_q$ is quasi-dipole latitude \cite{Richmond95}, $\phi_\mathrm{mlt}$ is the magnetic local time \cite{Laundal17}, $h$ [km] is the altitude, $R_E = 6371.2$~km is the Earth radius, $P_n^m(\sin \lambda_q)$ are Schmidt semi-normalized associated Legendre functions of degree $n$ and order $m$, and the spherical harmonic coefficients $g_n^m$ and $h_n^m$ are specific functions of solar wind speed, IMF $B_y$ and $B_z$, dipole tilt angle, and the F$_{10.7}$ solar flux index \cite{Laundal18b}. In the AMPS model, the truncation level of the double sum is set to $n\le45, m\le3$. The radial dependence of equation (\ref{eq:V_pol}) corresponds to a magnetic field of internal origin \cite{Chapman40}. It can be modeled by a sheet current $\mathbf{J}^V$ at some height $h_R < h$, where $\mathbf{J}^V = \mathbf{k}\times\nabla\Psi$, $\mathbf{k}$ is an upward unit vector, and
\begin{equation}
\Psi(\lambda_q, \phi_\mathrm{mlt}) = -\frac{R_E}{\mu_0} \sum_{n,m} \frac{2n+1}{n}\left( \frac{R_E}{R_E + h_R} \right)^{n+1}P_n^m(\sin\lambda_q)\left[g_n^m\cos m\phi_{\mathrm{mlt}} + h_n^m\sin m\phi_{\mathrm{mlt}}\right]\label{eq:psi},
\end{equation}
where $\mu_0$ is the vacuum permeability ($4\pi\cdot 10^{-7}$~H/m). Since equation (\ref{eq:V_pol}) refers to a magnetic field of internal origin, relative to altitudes above the ionospheric horizontal currents, it is not suitable for use below the ionosphere, where the ionospheric currents are external. Below the ionospheric currents, the magnetic potential field is often split in two parts, $\mathbf{B} = -\nabla V^e-\nabla V^i$, one ($V^e$) that corresponds to external sources (ionospheric currents) and another ($V^i$) that corresponds to currents below the Earth's surface, that are induced by the external currents. A spherical harmonic expansion of $V^i$ is exactly analogous to the expansion of $V$ in equation (\ref{eq:V_pol}), while the expansion of $V^e$ has a different radial dependence:
\begin{equation}
V^e(\lambda_q, \phi_\mathrm{mlt}, h) = R_E\sum_{n,m} P_n^m(\sin\lambda_q)\left(\frac{R_E +h}{R_E}\right)^n[a_n^m\cos(m\phi_\mathrm{mlt}) + b_n^m\sin(m\phi_\mathrm{mlt})] \label{eq:V_cb}.
\end{equation}
Just like the internal potential of equation (\ref{eq:V_pol}), the external potential can be modeled by a sheet current $\mathbf{J}^e$ at height $h_R > h$, where $\mathbf{J}^e = \mathbf{k}\times\nabla\Psi^e$ and
\begin{equation}
\Psi^e  = \frac{R_E}{\mu_0}\sum_{n, m} \frac{2n+1}{n+1}\left(\frac{R_E + h_R}{R_E}\right)^{n}P_n^m(\sin \lambda_q)[a_n^m\cos(m\phi_{\mathrm{mlt}}) + b_n^m\sin(m\phi_{\mathrm{mlt}})] \label{eq:equivalent_current}.
\end{equation}

We now make the assumption that the currents $\mathbf{J}^V$ = $\mathbf{J}^e$ = $\mathbf{J}_{df}$, so that $\Psi = \Psi^e + c$, where $c$ is an arbitrary constant. This equation allows us to express the spherical harmonic coefficients in equation (\ref{eq:equivalent_current}), $a_n^m$ and $b_n^m$, in terms of the AMPS model coefficients, $g_n^m$ and $h_n^m$:
\begin{equation}
\binom{a_n^m}{b_n^m} = -\frac{n+1}{n}\left(\frac{R_E}{R_E + h_R}\right)^{2n + 1}\binom{g_n^m}{h_n^m}.
\end{equation}
These expressions can be inserted in equation (\ref{eq:V_cb}), and thus we can calculate the magnetic field component of the external magnetic field below the ionosphere, $\mathbf{B}^e = -\nabla V^e$, in terms of AMPS model coefficients:
\begin{align}
B_\phi   &= -\frac{1}{\cos\lambda_q}\sum_{n,m}\left(\frac{R_E}{R_E + h_R}\right)^{2n+1}\frac{n+1}{n}P_n^m(\sin\lambda_q)m[g_n^m\sin(m\phi_\mathrm{mlt}) - h_n^m\cos(m\phi_\mathrm{mlt})] \label{eq:beg}\\
B_{\lambda_q}   &= \sum_{n,m}\left(\frac{R_E}{R_E + h_R}\right)^{2n+1}\frac{n+1}{n}\frac{dP_n^m(\sin\lambda_q)}{d\lambda_q}[g_n^m\cos(m\phi_\mathrm{mlt}) + h_n^m\sin(m\phi_\mathrm{mlt})] \label{eq:bng}\\{}
B_r   &=  \sum_{n,m}\left(\frac{R_E}{R_E + h_R}\right)^{2n+1}(n+1)P_n^m(\sin\lambda_q)[g_n^m\cos(m\phi_\mathrm{mlt}) + h_n^m\sin(m\phi_\mathrm{mlt})] \label{eq:bug}
\end{align}

The magnetic field components $B_\phi$, $B_{\lambda_q}$, and $B_r$, must be regarded as quasi-dipole components. For conversion to geographic east, west, and upward components, we use the quasi-dipole base vectors $\mathbf{g}_i$ and $\mathbf{f}_i$, which vary with the geometry of the main magnetic field \cite{Laundal17}:
\begin{equation}
\mathbf{B} = B_\phi F\mathbf{g}_1 +B_{\lambda_q} F\mathbf{g}_2 +B_r \mathbf{k}
\end{equation}
where $F = \mathbf{f}_1\times\mathbf{f}_2\cdot\mathbf{k}$. 

At least two critical approximations are involved in the approach described above: 1) The choice of 110~km altitude is somewhat arbitrary; lowering it would increase the ground magnetic field; 2) We ignore effects of ground-induced currents. Induced effects are in theory present in the AMPS field, but they will be stronger on ground, and will also be time-scale dependent \cite{Kuvshinov08}. 

In our analysis we derive a synthetic AL index from the AMPS model, and compare with measured AL values. We chose the AL index since it measures the westward electrojet, which typically peaks on the nightside in response to increases in magnetotail activity, most notably substorms. It is also comparatively straightforward to calculate: The synthetic AL values are calculated as the lower envelope of the H component of the AMPS model magnetic field at the AL station locations. The H component is the magnetic field along the horizontal part of Earth's main magnetic field (here we use the CHAOS model \cite{Finlay16}). An updated version (version 0103) of the AMPS model was used, defined and calculated as described in the original AMPS model paper\cite{Laundal18b}, but with a 43\% larger dataset which includes more recent {\it Swarm} data. In the updated version we have also corrected a programming error that damped the longitudinal variation in the toroidal magnetic field and associated magnetic field-aligned electric currents (FACs) of the original model. The correction leads to peak FACs that are significantly increased  (by typically $\sim20\%$) compared to the original model, however the integrated FACs differ by typically less than $\sim10\%$. The changes in poloidal magnetic field and associated horizontal divergence-free currents and ground magnetic field disturbances are small (typically $\sim1\%$). The new version is now integrated in the latest update (version 1.4.0) of the publicly available AMPS forward code, pyAMPS \cite{pyamps18}, which was used for this study. The model is also published as an ESA {\it Swarm} Level 2 data product. 

\subsection*{Results}
\begin{figure}
\includegraphics[width=\linewidth]{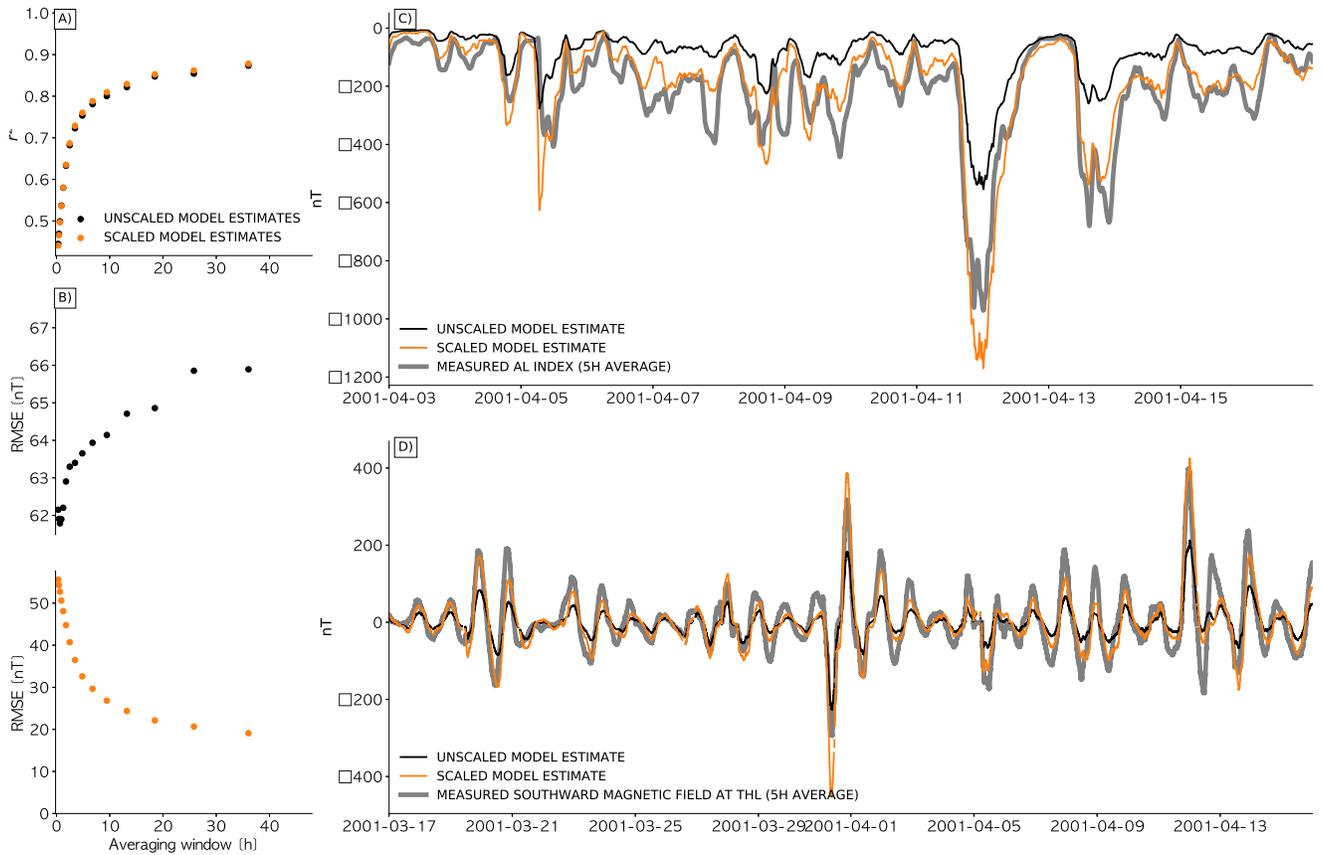}
\caption{Time scale dependence of AMPS model estimates of ground magnetic field perturbations. A) Pearson correlation coefficient squared between AMPS model predictions and measured values of the AL index, as function of averaging window. Black dots correspond to unscaled estimates, using the approach for calculating AMPS ground magnetic field outlined in the main text. The orange dots correspond to scaled estimates, using equation (\ref{eq:scale}). B) Root mean square error (RMSE) of AMPS AL estimates, in units of nanotesla [nT]. Black (orange) correspond to unscaled (scaled) AMPS values. C) Example time series of running 5~h mean AL index (grey) compared to unscaled (black) and scaled (orange) AMPS estimates. D) Example time series of 5~h running mean southward magnetic field disturbance at the Thule station, at approximately $85^\circ$ quasi-dipole latitude, compared with unscaled (black) and scaled (orange) AMPS estimates. }
\label{fig:amps}
\end{figure}

Figure \ref{fig:amps} summarizes the results of the comparison between AMPS model predictions and ground magnetic field measurements. To the left, we show the $r^2$ and RMSE between model and measured AL, in black. The RMSE is calculated using iterative reweighting with Huber weights, to reduce the effect of outliers. The coefficient of determination ($r^2$) increases with time scale, as expected. The increase in RMSE with time scale is more surprising. The model increasingly underestimates the data points as time scale increases. The reason for this is probably the time-scale dependent relationship between $\epsilon$ and $\Phi_N$: The AMPS model is based on relatively high time resolution data, and is thus scaled to fit the directly driven part of the magnetic disturbances. When compared with data on longer time scales, it correlates with both the loading and unloading parts of the disturbances, but the amplitudes still represent only one part. 

The time-scale dependence discussed in the previous sections suggest that the AMPS model estimates could be improved by including a time scale-dependent scale factor for the $\epsilon\mathbf{B}_\epsilon$ term in equation (\ref{eq:amps_decomposition}). We estimate such a scale factor using the model and measured AL time series from the period 1995 to 2018. For each time scale considered in the scatter plots in Figure \ref{fig:amps}, we find the scale factor that minimizes the Huber weighted root mean square model data mismatch. The resulting points, plotted against time scale $\tau$, are approximately sigmoidal, and we therefore estimate the $\tau$-dependent scaling factor as a logistic function plus a constant:
\begin{equation}
C(\tau) = 1.0 + 1.18(1 + e^{-0.01(\tau - 28)})^{-1} \label{eq:scale},
\end{equation}
with $\tau$ given in minutes. Multiplying the $\epsilon\mathbf{B}_\epsilon$ term in equation (\ref{eq:amps_decomposition}) by $C(\tau)$ results in the orange dots in Figures \ref{fig:amps}A and B. We see that the correlation is almost unaffected. This is expected, because the scaled version does not include any more information about the high-frequency component of nightside processes. However, the RMSE is significantly improved: At time scales of the order 24~h the RMSE is only 20~nT, and about 90\% of the variation is explained. 

Example time series are shown in Figure \ref{fig:amps}C-D. Figure \ref{fig:amps}C shows the AL index measurements averaged over 5~h in grey, and corresponding AMPS estimates in black and orange. The black curve is unscaled, and clearly underestimates the fluctuations in AL. The orange curve is scaled using $C(\tau)$ in equation (\ref{eq:scale}), and is much closer to the measurements. 

Figure \ref{fig:amps}D shows measurements of the ground magnetic field at the Thule station in Northern Greenland together with AMPS model estimates. The plot represents perturbations in the southward direction averaged over 5~h. The black curve represents unscaled model estimates, and the orange curve represents model estimates scaled using $C(\tau)$ in equation (\ref{eq:scale}). The close fit indicates that the scale factor derived using AL index comparisons can be applied to other ground magnetometers as well.

The scale factor $C(\tau)$ given by equation (\ref{eq:scale}) is 1.51 at small $\tau$. This may imply that the AMPS model ground perturbations underestimate the measurements at short time scales, even though the model was derived using high time resolution data: 1~Hz satellite measurements and $\tau=20$~min solar wind data. It is possible that this is because the AMPS model lacks the spatial resolution to see localized instantaneous features. It is also possible that this discrepancy is due to the neglect of induced effects in our calculations of ground magnetic field disturbances. The induction effect, which is likewise time scale dependent, also influences the scale factor $C(\tau)$ in equation (\ref{eq:scale}), such that $C(\tau)$ can not be interpreted as a pure effect of the time-scale dependence between dayside and nightside reconnection rates.

\section*{Estimating nightside reconnection rate using geomagnetic indices and solar wind data}\label{sec:phin}

\begin{figure}
\includegraphics[width=\linewidth]{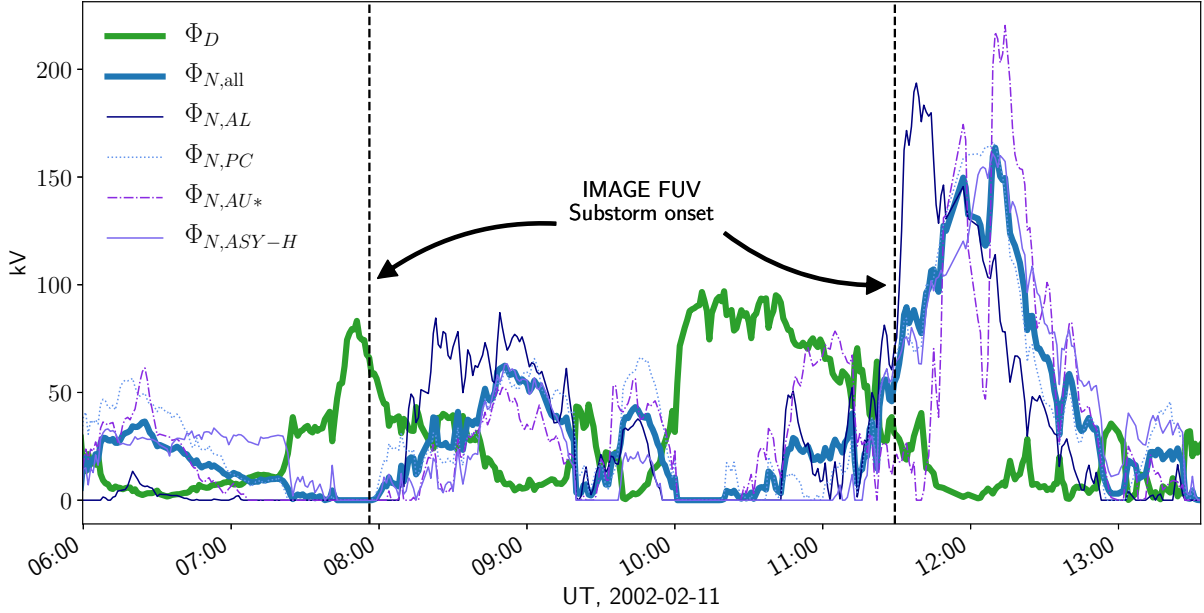}
\caption{Example estimate from November 2002 of nightside reconnection rate (blue), using equation (\ref{eq:phin}). The thin lines represent $\Phi_N$ estimates based on the four individual indices, and the thick line represents their mean value. The green line shows the dayside reconnection rate, according Milan's solar wind magnetosphere coupling function\cite{Milan12}. The period shown contained two substorm onsets according to the Frey IMAGE FUV based list\cite{Frey04}. The onset times are marked by dashed vertical bars.}
\label{fig:phin}
\end{figure}

We have shown how measurements of $y$ and $\Phi_D$ can be used to estimate the nightside reconnection rate $\Phi_N$ on any time scale via equation (\ref{eq:phin}). Figure \ref{fig:phin} shows (in thick blue) an example time series $\Phi_{N}$, calculated as the average of four different estimates obtained with the AL, PC, AU, and ASY-H indices. The thin blue lines show the $\Phi_N$ estimates calculated using the four individual indices. The green curve shows $\Phi_D$, as estimated with the Milan\cite{Milan12} coupling function. Both time series are shown with 1~min resolution. The two vertical bars denote the time of substorm onsets observed in global images of the aurora produced by the Far Ultra Violet (FUV) instrument on the IMAGE satellite \cite{Frey04}. We see that both substorm onsets are preceded by a period of strong dayside reconnection (growth phase), and followed by a period of strong tail reconnection (substorm expansion).

Figure \ref{fig:phin} shows only a small extract of a time series calculated for the years 1995 through 2014, containing 8.3~million 1~min values. During these years the Pearson correlation coefficient between $\Phi_N$ and $\Phi_D$ is 0.2, which means that only $4\%$ of the variation in $\Phi_N$ can be explained by $\Phi_D$ for a time scale $\tau=1$~min. This tells us that $\Phi_N$, estimated from equation (\ref{eq:phin}), contains information which is not in $\Phi_D$. Since the calculation can be done whenever solar wind data is available to calculate $\Phi_D$, and since geomagnetic indicies typically have very few data gaps, this may be a useful parameter to include in future empirical models of ionospheric electrodynamics. 

The correlation $r$ ($r^2$) between $\Phi_D$ and $\Phi_N$ at $\tau = 8$~days is 0.85 (0.72). If equation (\ref{eq:phin}) and Milan's coupling function\cite{Milan12} gave perfect estimates of $\Phi_N$ and $\Phi_D$, we would instead find $r\approx1$ at this time scale. The difference indicates that there is a need for closer scrutiny of the $\Phi_N$ estimates. In future work we plan to carry out more detailed calibration and validation of our estimates of $\Phi_N$ by comparing with other independent estimates. For example, previous estimates of $\Phi_N$ have been based on observations of the polar cap together with the Milan\cite{Milan12} coupling function. The polar cap is the area threaded by open magnetic field lines. By observing this area over some time, the change in open flux $F$ can be calculated, and the nightside reconnection rate calculated as $\Phi_N = \Phi_D - dF/dt$. This technique is good because it uses the definitions of $\Phi_N$ and $\Phi_D$ directly, instead of the indirect approach outlined here. A major disadvantage is that $F$ must be measured precisely and continuously, which is very difficult to do. So far it has only been done in limited time intervals, for example periods when the IMAGE satellite could observe the whole auroral oval \cite{Hubert06,Laundal10a,Ohma18}. It can in principle also be done by inferring the polar cap boundary position from magnetometer measurements at low Earth orbit \cite{Milan17}.

\section*{Discussion}
The two-step response of the magnetosphere-ionosphere system to changes in the solar wind has been known for decades, and it is an integral part of the expanding contracting polar cap paradigm \cite{Cowley92,Milan07}. The contribution of the present study is to quantify the time-scale dependent statistical relationship between nightside and dayside reconnection rates, as well as its effect on solar wind based regression models of ionospheric electrodynamic parameters; and to use this relationship to derive equation (\ref{eq:phin}), which following more detailed calibration and validation may prove useful as a monitor of the nightside reconnection rate. 

We have shown that solar wind-based regression models of ionospheric electrodynamics are time scale dependent: Model parameters depend on the time resolution of the input data. We have demonstrated this with simple linear regression models of the AL, AU, PC, and ASY-H indices. We have also shown that this applies to the more complex AMPS model. By comparing AMPS model predictions with measured values of the AL index, we have derived the time-scale dependent correction factor $C(\tau)$ given in eq. (\ref{eq:scale}), which can be used with the AMPS model when estimating time-averaged magnetic field disturbances. The Python AMPS forward code, pyAMPS \cite{pyamps18}, includes support for this scaling in the get\_B\_ground function, via the epsilon\_multiplier keyword argument.

We have also presented a simple model for the nightside reconnection rate, equation (\ref{eq:phin}), which depends on geomagnetic indices and on the dayside reconnection rate. The latter can be estimated from solar wind parameters \cite{Milan12}. This derivation assumes that the geomagnetic indices we have considered are proportional to the cross polar cap potential (i.e., they obey equation (\ref{eq:ideal_model})), and takes into account that nightside and dayside reconnection rates balance over time. The high correlation between linear models and corresponding indices on long time scales, seen in Figure \ref{fig:param_model_stats}, indicate that this assumption is reasonable. However, it is not perfect, since that would imply that all the indices are perfectly correlated with each other. In this study we have not attempted to assess which index, when used in equation (\ref{eq:phin}), gives the most accurate estimate of nightside reconnection rate. By comparing different realizations of this equation, using combinations of indices with independent estimates of nightside reconnection, it should be possible to derive an empirical function that describes the ''unloading'' response, similar to how empirical solar wind-magnetosphere coupling functions describe ''loading'', or direct driving. Such a function will be the topic of a future study. The inclusion of an unloading parameter in physics-based empirical models, like the AMPS model, may help to determine the role of individual driving processes in controlling the ionospheric electrodynamics.

\section*{Acknowledgements}

The study was funded by the Research Council of Norway/CoE under contract 223252/F50. CCF and CK were supported by the European Research Council (ERC) under the European Union's Horizon 2020 research and innovation programme (grant agreement No. 772561). The AMPS model coefficients are available from ESA through https://swarm-diss.eo.esa.int, and Python forward code can be found at https://github.com/klaundal/pyAMPS. The IMF, solar wind and magnetic index data were provided through OMNIWeb by the Space Physics Data Facility(SPDF), and downloaded from ftp://spdf.gsfc. nasa.gov/pub/data/omni/highresomni/. The Qaanaq (THL) magnetometer is operated by the National Space Institute, Technical University of Denmark (DTU Space) and data can be obtained via the Troms\o{} Geophysical Observatory website, http://flux.phys.uit.no/geomag.html or the INTERMAGNET website, http://www.intermagnet.org.  INTERMAGNET is thanked for promoting high standards of magnetic observatory practice.

\section*{Author contributions statement}
K.M.L. conceived of and implemented the analysis, and wrote the manuscript. J.P.R., S.M.H., T.M., A.O., N.\O{}, P.A.R.T., C.C.F., and C.K. contributed to the analysis. All authors reviewed the manuscript.

\end{document}